\documentclass[a4paper,11pt]{article}
\usepackage{color, colortbl}
\definecolor{Gray}{gray}{0.9}
\usepackage{jinstpub} 

\usepackage{dcolumn}
\usepackage{lineno}

\title{Considerations for a TeV Collider Based On Dielectric Laser Accelerators}

\author[a,1]{R. J. England,\note{Corresponding author.}}
\author[b]{U. Niedermayer,}
\author[c]{L. Sch\"achter,}
\author[d]{T. Hughes,}
\author[e]{P. Musumeci,}
\author[f]{R. K. Li,}
\author[g]{and W. D. Kimura}

\affiliation[a]{SLAC National Accelerator Laboratory,2575 Sand Hill Rd., Menlo Park, CA, USA}
\affiliation[b]{Technical University of Darmstadt, Institute for Accelerator Science and Electromagnetic Fields, Schlossgartenstrasse 8, 64289 Darmstadt, Germany}
\affiliation[c]{Technion - Israel Institute of Technology, Haifa, Israel}
\affiliation[d]{Flexcompute, Inc., 910 Foulk Rd, Suite 201, Wilmington DE, USA}
\affiliation[e]{University of California Los Angeles, 405 Hilgard, Los Angeles, CA, USA}
\affiliation[f]{Tsinghua University, 30 Shuangqing Rd, Beijing, China}
\affiliation[g]{STI Optronics Inc., 1809 130th Ave NE Suite 118, Bellevue, WA, USA}

\emailAdd{england@slac.stanford.edu}

\abstract{
Particle acceleration in dielectric microstructures powered by infrared lasers, or ``dielectric laser acceleration" (DLA), is a promising area of advanced accelerator research with the potential to enable more affordable and higher-gradient accelerators for energy frontier science and a variety of other applications. DLA leverages well-established industrial fabrication capabilities and the commercial availability of tabletop lasers to reduce cost, with axial accelerating fields in the GV/m range. Desirable luminosities would be obtained by operating with very low charge per bunch but at extremely high repetition rates. And as a consequence of its unique operating parameter regime, coupling of the laser to the accelerator can potentially be in the 50\% range and with low beamstrahlung energy loss due at the interaction point, making DLA a promising approach for a future multi-TeV linear collider.
}


\proceeding{Contribution for Beam Dynamics Newsletter \#83 - Special Issue on Challenges in Advanced Accelerator Concepts\\
  February 2022\\}

\begin{document}
\maketitle
\flushbottom

\section{Introduction}
\label{sec:intro}

Modern state-of-the-art particle accelerators operate at microwave frequencies and use metallic cavities to confine electromagnetic modes with axial acceleration forces.  The achievable fields in these devices are ultimately limited by the electrical breakdown of the metallic surfaces to accelerating fields of order 10 to 50 MV/m.  However, the basic technologies employed in modern accelerators (metal cavities powered by microwave klystrons) are over 50 years old.  Even particle accelerators with modest particle energies of a few hundred MeV are large and expensive devices accessible mainly to government laboratories.  The largest accelerators used for high energy physics have construction costs in the billions of dollars and occupy many kilometers of real estate.  Constraints on the size and cost of accelerators have inspired a variety of advanced acceleration concepts for making smaller and more affordable particle accelerators.

The use of lasers as an acceleration mechanism is particularly attractive in this regard, due to the intense electric fields they can generate combined with the fact that the solid state laser market has been driven by extensive industrial and university use toward higher power and lower cost over the last 20 years. Metallic surfaces have high ohmic loss and low breakdown limits at optical frequencies, making them generally undesirable as confining structures for laser-powered acceleration.  Dielectrics and semiconductor materials, on the other hand, have damage limits corresponding to acceleration fields in the 1 to 10 GV/m range, which is orders of magnitude larger than conventional accelerators. Such materials are also amenable to rapid and inexpensive CMOS and MEMS fabrication methods developed by the integrated circuit industry. These technological developments over the last two decades, combined with new concepts for efficient field confinement using optical waveguides and photonic crystals, and the first demonstration experiments of near-field structure-based laser acceleration conducted within the last few years, have set the stage for making integrated laser-driven micro-accelerators or ``dielectric laser accelerators" (DLA) for a variety of real-world applications. The current state-of-the-art in DLAs includes accelerating gradients approaching 1 GV/m \cite{wootton_demonstration_2016, cesar_nonlinear_2018} and energy gains on the order of 300 keV \cite{peralta:2013,cesar:pft:2018}. DLA has the potential to provide high efficiency accelerators operating at very high repetition rates, with bunch formats (charge, beam sizes, emittances, time duration and temporal separation) significantly different than what is commonly available in conventional accelerator facilities.  

\begin{figure}
\begin{center}
 \includegraphics[height=.35\textheight]{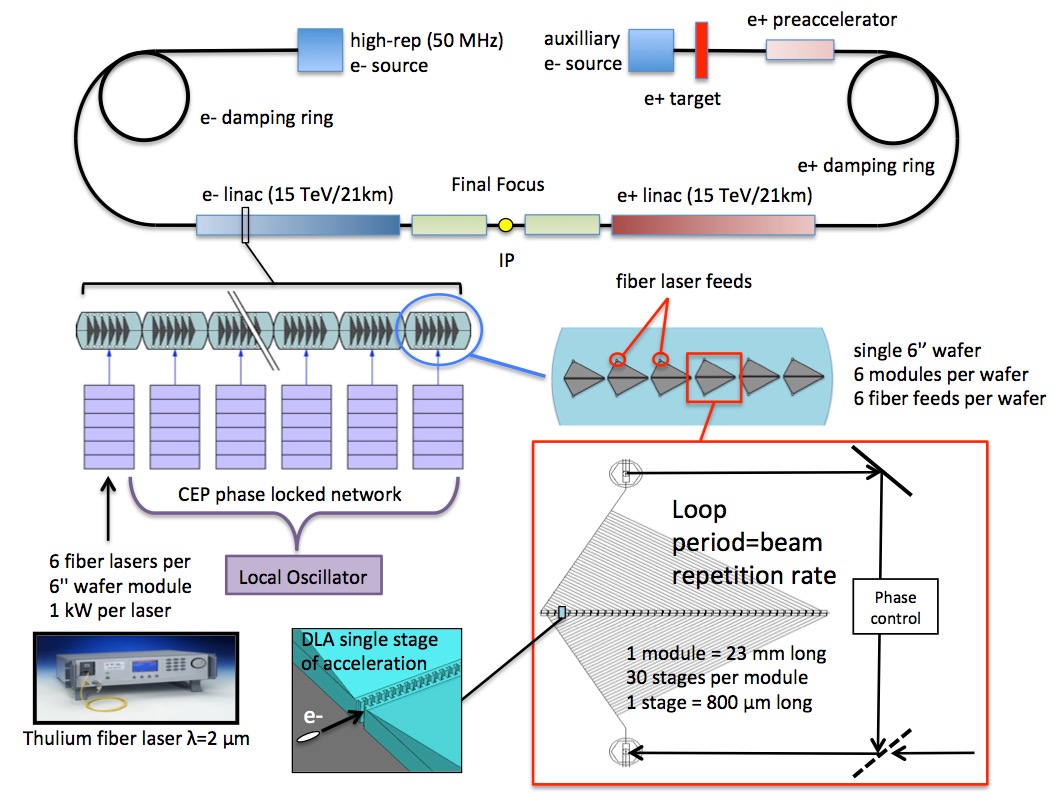}
\caption{Conceptual schematic of a 30 TeV DLA e+ e- collider driven by a carrier envelope phase locked network of energy-efficient solid-state fiber lasers at 20 MHz repetition rate. Laser power is distributed by photonic waveguides to a sequence of dielectric accelerating, focusing, and steering elements co-fabricated on 6-inch wafers which are aligned and stabilized using mechanical and thermal active feedback systems.}
\label{collider}
\end{center}
 \end{figure}

\section{Roadmap for a DLA Collider}
\label{sec:anar}

A future DLA-based linear collider, schematically illustrated in Fig.~\ref{collider}, will require the development of high-gradient accelerator structures as well as suitable diagnostics and beam manipulation techniques, including compatible small-footprint deflectors, focusing elements, and beam position monitors (BPMs).  Key developments in these areas have been made within the last 5 years, including the demonstration of high average gradients (300--850 MeV/m) with speed-of-light synchronous acceleration in laser-driven dielectric microstructures \cite{peralta:2013, wootton_demonstration_2016, cesar_nonlinear_2018}, non-relativistic acceleration with gradients up to 350 MV/m \cite{breuer_laser-based_2013,leedle_dielectric_2015}, and development of preliminary design concepts for compatible photonic components and power distribution networks \cite{hughes:chip:2018,mcneur_elements_2018}. The power distribution scheme is then envisioned as a fiber-to-chip coupler that brings a pulse from an external fiber laser onto the integrated chip, distributes it between multiple structures via on-chip waveguide power splitters, and then recombines the spent laser pulse and extracts it from the chip via a mirror-image fiber output coupler \cite{colby:2011}, after which the power is either dumped, or for optimal efficiency, recycled \cite{siemann:2004}.  Maintaining phase synchronicity of the laser pulse and the accelerated electrons between many separately fed structures could be accomplished by fabricating the requisite phase delays into the lengths of the waveguide feeds and employing the use of active feedback systems.  

The DLA mechanism is also equally suitable for accelerating both electrons and positrons. Example machine parameters for a DLA collider have been outlined in the Snowmass 2013 report and several other references \cite{colby:2011,dla:2011,snowmass:2013,england:rmp2014}.  In these example parameter studies, DLA meets desired luminosities with reasonable power consumption and with low beamstrahlung energy loss \cite{beambeam:2021}.

\begin{figure}
\begin{center}
 \includegraphics[height=.35\textheight]{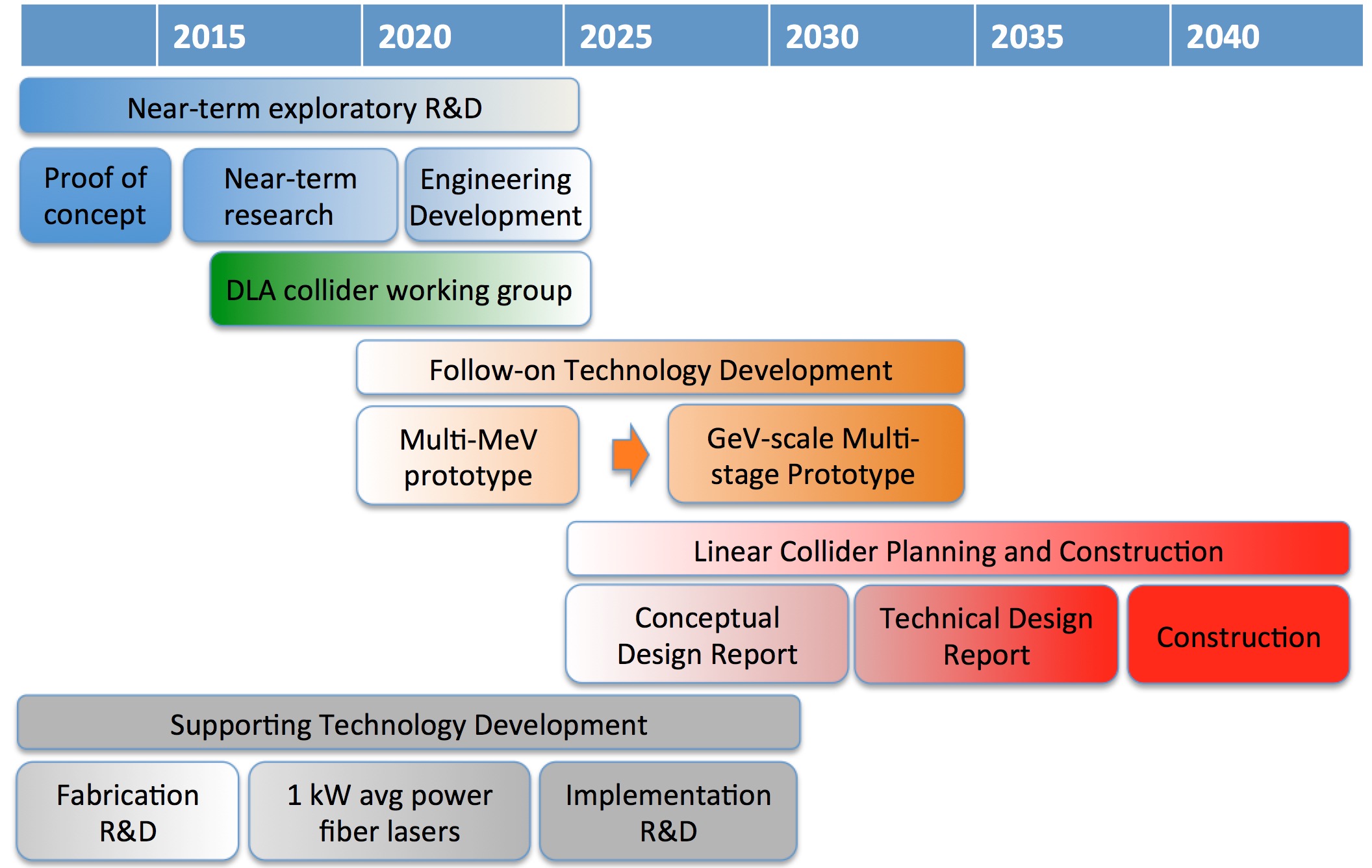}
\caption{Thirty-year roadmap for a DLA collider, reproduced from the ANAR 2017 Working Group 4 Report \cite{anar:2017}.}
\label{30Year}
\end{center}
 \end{figure}

The Advanced and Novel Accelerators for High Energy Physics Roadmap Workshop (ANAR) was held at CERN in June 2017, with the goal of identifying promising advanced accelerator technologies and establishing an international scientific and strategic roadmap toward a future high energy physics collider \cite{anar:2017}.  Four concepts were considered:  laser-driven plasma wake field acceleration (LWFA), beam-driven plasma wake field acceleration (PWFA), structure-based wake field acceleration (SWFA), and dielectric laser acceleration (DLA). Dedicated working groups were convened to study each concept.  The working group on DLA produced a roadmap to a DLA based collider on a 30 year time scale, as shown in Fig.~\ref{30Year}. Compact multi-MeV DLA systems for industrial and scientific use are expected on a 10 year time scale.  A dedicated GeV-scale multi-stage prototype system is recommended within 20 years to demonstrate energy scaling over many meters with efficiency and beam quality suitable for HEP applications. 

The DLA working group evaluated current state of the art in the field and identified both significant advantages as well as technical challenges of the DLA approach as a future collider technology. Key advantages include the fact that the acceleration occurs in vacuum within a fixed electromagnetic device, that the acceleration mechanism works equally well for both electrons and positrons, and that the approach is readily amenable to nanometrically precise alignment and optical stabilization of multiple stages.  Complex integrated photonic systems have been shown to provide phase-stable operation for time periods of order days \cite{hulme:2014,xiang:2016}, and nanometric alignment of optical components over kilometer-scale distances has been well established by the LIGO project with stability of 0.1 nm Hz$^{-1/2}$ \cite{ligo:stability}.  In addition, the low-charge and high-repetition-rate particle bunch format inherent to the DLA scheme would provide a very clean crossing at the interaction point of a multi-TeV collider, with estimated beamstrahlung losses in the single percent range, as compared with 10s of percents for more conventional accelerators \cite{beambeam:2021}. 

Technical challenges were prioritized from High to Low, with higher priority items being addressed earlier in the timeline. Supporting technologies, including high average power solid state lasers and precise nanofabrication methods, were deemed low priority, since the current state of the art in these areas is already at or near required specifications. Detailed considerations of the final focus design and beam collimation were also deemed lower priority:  Due to the very low charge and low emittance beams that a DLA accelerator would intrinsically provide, existing approaches for more conventional accelerators would already be over-engineered for the DLA scenario and could thus be directly applied or perhaps even made more compact. The highest priority challenges identified by the ANAR working group largely pertain to the transport of high average beam currents in the relatively narrow (micron-scale) apertures of nanostructured devices.  These include effects such as beam breakup instability, charging, radiation damage, and beam halo formation, which may be less relevant at low beam powers and beam energies, but can be highly detrimental in a collider scenario.  To these ends, it was recommended to establish a core working group on DLA that would oversee the strawman collider design and motivate these feasibility studies.

\section{Test Facilities for DLA}
\label{sec:facilities}
{\hskip 0.13in}
Dielectric laser accelerators have a lot of ground to cover to be competitive with other advanced accelerator techniques which have already demonstrated gradients in excess of 50 GV/m and multi-GeV energy gains. For these reasons, it is important for a beam test facility to match the unique characteristics of the DLA accelerators. For example, very high repetition rate electron sources combined with compact, efficient, and relatively low energy ($\mu$J to mJ class) lasers will enable testing of linear-collider relevant concepts such as beam loading and wall-plug efficiency. In this regard we note a general trend in DLA research towards longer wavelengths.  This is motivated in part by a desire to increase the phase space acceptance of the accelerator, which scales with the wavelength both in longitudinal and transverse dimensions. Therefore availability of suitable laser driver pulses in the mid-infrared will be needed. 

Compatibility with DLA's unique features requires high brightness beamlines equipped with diagnostics suitable for the measurement of ultralow (sub-pC to few-fC) bunch charges and ultralow (< 1 nm-rad) normalized emittance. Current relativistic DLA experiments have mostly taken place at low repetition rate facilities such as the next linear collider test accelerator (NLCTA) at SLAC, the Athos beam line at SwissFEL, the SINBAD facility at DESY, and the Pegasus facility at UCLA \cite{peralta:2013,wootton_demonstration_2016,cesar_nonlinear_2018}. Such facilities are sufficient for initial proof-of-principle experiments, but should in future be coupled with the novel electron sources being developed for high repetition rate free-electron lasers such as superconducting and very high-frequency RF guns. 

A significant body of exploratory research has also been carried out using refurbished electron microscope columns \cite{breuer:2013,leedle:2015,leedle_dielectric_2015,mcneur_elements_2018}.  However, it should be noted that the brightness of these sources is not sufficient to efficiently couple the beam to the small phase space acceptances typical of DLAs. Furthermore, an analysis of coupling efficiency shows that at least mildly relativistic electron energies are strongly favored to maximize the accelerating gradients in the first stages of acceleration.  Another very important characteristic will be the availability of a complete suite of optical diagnostics to monitor performance and provide active feedback of the laser illumination of the DLA structures.  With all these elements available, the logical progression of DLA experiments to demonstrate suitability for high energy physics applications includes bunching, beam control over longer distances, staging, wakefield mitigation, beam halo, emittance preservation, and efficient energy transfer. For demonstrating multi-staged DLA accelerators, ultra-low emittance particle sources need to be developed and combined with DLA devices to make compact injectors. Ongoing development of DLA prototype integrated systems will provide a pathway for scaling of this technology to high energy (MeV to GeV) and to beam brightness of interest both for high energy physics and for a host of other applications, as discussed in Ref.~\cite{england:review:2016}.

It is likely that the path to a multi-TeV linear collider will require a multi-stage GeV-scale prototype to demonstrate the feasibility of the candidate collider technology or technologies to confirm gradient, emittance control, and wall-plug to beam efficiency, and to validate the fabrication cost model \cite{P5:2014}. Prior to building a large facility based upon a cutting-edge concept, a demonstration system of intermediate scale is well advised.  The primary purpose of such a demonstration system would be to incorporate interrelated technologies developed under a prior sequence of R\&D steps in order to identify and address new challenges arising from the integration of these components.  Ideally, such a demonstration system would combine all or most of the technological sub-units needed to build a larger-scale system, and would simultaneously possess utility in its own right as a compelling scientific tool.  We envision such a DLA-based demonstration system to consist of a sequence of wafer-scale modules, which each incorporate of order tens of single-stage acceleration sections individually driven by on-chip fiber or SOI type guided wave systems for directing laser light and phasing it in sync with the passing speed-of-light particle beam, as described in Refs. \cite{england:rmp2014,colby:2011,dla:2011,snowmass:2013}.  Such a system would illustrate:  (1) integration of the DLA concept with compatible MeV particle sources with nanometric beam emittance and attosecond particle bunch durations, (2) implementation of an accelerator architecture with a pathway to TeV beam energies, (3) carrier envelope phase-lock synchronization of multiple lasers and correct phasing and delivery of laser light over multiple acceleration stages, (4) beam alignment and steering between wafer-scale modules using interferometric techniques combined with feedback, and (5) efficient power handling and heat dissipation.  Due to the high potential cost of such a multi-stage prototype, cost sharing may be possible if the prototype also serves a secondary purpose, such as driving a future light source.

\section{Scaling Considerations for a Linear Collider}
\label{sec:parameters}
{\hskip 0.13in}
To reach 30 TeV center-of-mass energies, a next generation lepton collider based on traditional RF microwave technology would need to be over 100 km in length and would likely cost tens of billions of dollars to build.  Due to the inverse scaling of the interaction cross section with energy, the required luminosity for such a machine would be as much as 100 times higher than proposed 1 to 3 TeV machines (ILC and CLIC), producing a luminosity goal of order $10^{36}$ cm$^{-2}$s$^{-1}$.  In attempting to meet these requirements in a smaller cost/size footprint using advanced acceleration schemes, the increased beam energy spread from radiative loss during beam-beam interaction (beamstrahlung) at the interaction point becomes a pressing concern.  Since the beamstrahlung parameter is proportional to the number of interacting particles, a straightforward approach to reducing it is to use low-charge bunches, with the resulting quadratic decrease in luminosity compensated by higher repetition rates.  This is a natural operating regime for the DLA scheme, with the requisite average laser power (>100 MW) and high (>10 MHz) repetition rates to be provided by modern fiber lasers.

Strawman parameters for the 250 GeV and 3 TeV cases have been previously reported \cite{england:rmp2014,rast:2016}, and these are reproduced in Table~\ref{parameters}.  To scale this scenario to 30 TeV we note that the total wall-plug power is proportional to the beam power $P_\text{wall} = P_\text{beam}/\eta$, where $\eta$ is the wall-plug efficiency and the beam power (of both beams together) is $P_\text{beam} = E_\text{cm} n N f_\text{rep}$, where $n$ is the number of bunches per train, $N$ the number of electrons per optical microbunch and $E_\text{cm}$ the center-of-mass energy.  The geometric luminosity scales as
\begin{equation}
\mathcal{L} = \frac { (n N)^2 f_\text{rep}} {4 \pi \sigma_x \sigma_y} = \chi E_\text{cm}^2 ,
\label{eq:luminosity}
\end{equation}
where $\chi = 2.3 \times 10^{33} \text{cm}^{-2} \text{s}^{-1} \text{TeV}^{-2}$ is a scaling constant \cite{king:2000}. We note that the luminosty here scales as $(n N)^2$ rather than $n N^2$ because it is assumed that entire bunch trains (each a single laser pulse in duration) collide at the IP.  For purposes of calculating the disruption parameter, luminosity enhancement, and beamstrahlung energy loss, it is further assumed that the microbunch structure is smeared out prior to the IP, giving a luminosity enhancement of approximately 10.  Hence the required particle flux scales as 
\begin{equation}
n N f_\text{rep} = E_\text{cm} \sqrt{4 \pi \sigma_x \sigma_y \chi f_\text{rep}} .
\label{eq:flux}
\end{equation}
Combining these relations we obtain the following scaling for wall-plug power with center-of-mass energy:
\begin{equation}
P_\text{wall} = \eta^{-1} E_\text{cm}^2 \sqrt{ 4 \pi \sigma_x \sigma_y \chi f_\text{rep}} .
\label{eq:wallplug}
\end{equation}
Consequently, if the center of mass energy is increased by a factor of 10 (from 3 TeV to 30 TeV), then for similar repetition rate and efficiency, the wall-plug power will increase by a factor of 100, as reflected in Table~\ref{parameters}.  At the same time, Eq.~\eqref{eq:flux} requires a 10 times increase in average beam current.  Due to the scaling in Eq.~\eqref{eq:flux} with $f_\text{rep}$ there is a tradeoff between charge and repetition rate. However, since the bunch charge $N$ and laser pulse duration are constrained by the efficiency, gradient, and space charge arguments of Section \ref{structures}, a potential solution is to incorporate 10 parallel DLA beamlines in a matrixed configuration such as that of the 2D honeycomb DLA of Fig.~\ref{structures}(b). The physics of the beam recombination mechanism at the IP requires further study, but for the purposes of Table~\ref{parameters} we assume a linear emittance scaling with number of parallel beamlines. In these examples, DLA meets the desired luminosity, and with a small (few percent) beamstrahlung energy loss.  Although the numbers in Table~\ref{parameters} are merely projections used for illustrative purposes, they highlight the fact that due to its unique operating regime, DLA is poised as a promising technology for future collider applications.

\begin{table}[]
\caption{DLA Strawman Parameters at 250 GeV, 3 TeV, and 30 TeV Energies}
\label{parameters}
\begin{center}
\begin{tabular}{lllll}
\hline
Parameter                 & Units & 250 GeV & 3 TeV   & 30 TeV  \\ \hline
Center of Mass Energy     & GeV   & 250     & 3000    & 30000   \\
Bunch charge              & e     & 3.8e4   & 3.0e4   & 3.0e4   \\ 
\# Bunches/train          & \#    & 159     & 159     & 159     \\ 
\# Parallel Beamlines     & \#    & 1       & 1       & 10      \\
Train Repetition Rate     & MHz   & 20      & 20      & 20      \\
Final Bunch Train Length  & ps    & 1.06    & 1.06    & 1.06    \\
Single Bunch Length       & $\mu$m    & 2.8e-3   & 2.8e-3   & 2.8e-3   \\
Drive Wavelength          & $\mu$m    & 2       & 2       & 2       \\
IP X Emittance (Norm)     & nm    & 0.1     & 0.1     & 1       \\
IP Y Emittance (Norm)     & nm    & 0.1     & 0.1     & 1       \\
IP X Spot Size            & nm    & 2       & 1       & 1       \\
IP Y Spot Size            & nm    & 2       & 1       & 1       \\
Beamstrahlung Energy Loss & \%    & 0.6     & 1.0     & 2.6     \\
Length of Beam Delivery   & m     & 2321    & 2304    & 2304    \\
Effective L*              & m     & 5       & 5       & 5       \\
Total Length              & km    & 5.0     & 8.4     & 42.1    \\
Geometric Luminosity      & cm$^{-2}$s$^{-1}$  & 1.46e33 & 3.63e33 & 3.63e35 \\
Enhanced Luminosity       & cm$^{-2}$s$^{-1}$  & 1.84e34 & 3.19e34 & 3.19e36 \\
Beam Power (per beam)     & MW    & 2.4     & 22.9    & 2292.5  \\
Total Wallplug Power      & MW    & 88.1    & 360.3   & 30487.4 \\
Wallplug Efficiency       & \%    & 5.5     & 12.7    & 15.0    \\ \hline
\end{tabular}
\end{center}
\end{table}


\section{Elements of a Future Linear Collider}
\label{sec:machine}

\subsection{Electron and Positron Sources}
\label{sec:sources}
{\hskip 0.13in}
Considerable effort at the university level has been directed towards development of DLA-compatible compact electron sources based on laser-assisted field emission from nanotips, which have been demonstrated to produce electron beams of unprecedented brightness. An alternative approach outlined in Ref.~\cite{schachter:kimura:2017} combines static and laser fields to extract electrons from carbon nano-tubes via field emission. However, comparable techniques for generating high-brightness positron beams in a compact footprint have not been identified.  While these miniaturized particle sources are well adapted to the DLA approach and are highly desirable for a variety of applications, they are less critical at the size and cost scales of a collider facility.  Consequently, a possible solution would be to use a high repetition rate cryogenic RF photoinjector as an electron source, with a separate DLA linac for positron production at a target followed by a damping ring.  A feasibility study is needed to understand how to adapt such a source to meet linear collider luminosity requirements with a DLA bunch format.

However, high gradient, high energy superconducting radio-frequency (SRF) guns operated in continuous wave (CW) mode are promising candidates for delivering relevant beams for DLA-based linear colliders. SRF guns have demonstrated reliable operation at 10 MV/m gradient, and there are active R\&D efforts to improve the gradient to >20 MV/m and even 40 MV/m. The CW operation allows in principle that each RF bucket be filled with a photoelectron bunch up to the resonant frequency of the cavity, which is typically 1.3 GHz for elliptical geometry guns and 100-200 MHz for quarter-wave resonator type guns \cite{arnold:2011}. The clean vacuum environment inside SRF guns also potentially allow advanced photocathodes to be used, while care must be taken to avoid contamination of the SRF cavity surface by nanoparticles from the cathode. Preliminary simulations show that it is possible to deliver 10 fC, 1 ps, 1.0 nm-rad emittance, 2 MeV electron beams from a 2 $\mu$m RMS laser spot on the cathode with 0.2 mm-mrad/mm RMS intrinsic emittance in a 20 MV/m, 200 MHz quarter-wave resonator type SRF gun. 


\subsection{Accelerating structures}
\label{sec:structures}
{\hskip 0.13in}
Use of lasers to accelerate charged particles in material structures has been a topic of considerable interest since shortly after the optical laser was invented in the early 1960s.  Early concepts proposed using lasers to accelerate particles by operating known radiative processes in reverse, including the inverse Cherenkov accelerator \cite{shimoda:1962} and the inverse Smith-Purcell accelerator \cite{takeda:1968, palmer:1980}. In the latter category, it was proposed by Palmer and Kroll to use the eigenmodes of an open planar grating to accelerate electrons \cite{palmer:1982,kroll:1985}. The exponential decay of the fields from the surface of such structures poses challenges for focusing and confinement of the particles, prompting schemes to solve this using (for example) a series of gratings successively rotated by 180 degrees about the acceleration axis \cite{pickup:1985}. Energy modulation of relativistic electrons has also been observed in a laser field truncated by a thin downstream metallic film \cite{leap:2005,sears:2008}. 

These early ideas and experiments laid the ground work for efficient phased laser acceleration.  However the interaction mechanism used to accelerate the particles in the experiments of Refs. \cite{takeda:1968, palmer:1980,leap:2005,sears:2008} is a relatively weak effect, requiring laser operation at fluences above the damage limit of the metal surface.  This points to the need to use materials with high damage limits combined with acceleration mechanisms that are more efficient.  Due to these considerations combined with the fact that metallic surfaces suffer high ohmic losses and low damage threshold limits at optical wavelengths, a shift in focus has occurred towards photonic structures made of dielectric materials and incorporating new technologies such as photonic crystals and meta-surfaces \cite{rosing:1990,lin:2001,cowan:2003,mizrahi:2004,schachter:2004,naranjo:2012,scheuer:2014}. The quasi-planar 1D type of geometry illustrated in Fig.~\ref{structures}(a) is simpler to fabricate and its wide aspect ratio helps in improving charge transmission, making demonstration experiments simpler.  This has led to structures of type (a) being the first to be successfully fabricated and undergo demonstration experiments \cite{peralta:2013,mcneur:2012}.  Due to the stringent requirements of a linear collider on beam quality and luminosity, we outline below various key constraints on DLA structure design and beam operation which inform the strawman collider scenario of Section \ref{sec:parameters}.  

\subsubsection{Loaded Gradient and Efficiency}
For the case of a 30 TeV collider, the average power carried by the electron beam is of order 0.5 GW. In order to get this power into the electron beam, we need twice this power (assuming 50\% laser-to-electron coupling efficiency) in the laser beam. It has been shown \cite{hanuka:single:2018} that maximum efficiency does not occur for the same parameters as maximum loaded gradient. Therefore we must either operate at maximum efficiency and compromise the gradient (increasing the required length of the accelerator) or operate at the maximum loaded gradient and compromise the efficiency and thus running into severe problems of wall-plug power consumption. Several possible regimes of operation have been analyzed in Ref.~\cite{hanuka:regimes:2018}.  For our considerations in Section \ref{sec:parameters}, we take as a conservative number a loaded gradient of 1 GeV/m, and a multi-bunch laser-electron coupling efficiency of 40\%. By microbunching the beam, the coupling efficiency of the axial laser field to the particles in a DLA can in principle be as high as 60\% \cite{siemann:2004,na:2005}.  Combined with recent advances in power efficiencies of solid state lasers, which now exceed 30\% \cite{moulton:2009} and designs for near 100\% power coupling of laser power into a DLA structure \cite{wu:2014}, estimates of wall-plug power efficiency for a DLA based system are in the range of 10--12\%, which is comparable to more conventional approaches \cite{england:rmp2014}.  

\subsubsection{Single Mode Operation}
\begin{figure}
\begin{center}
 \includegraphics[height=.18\textheight]{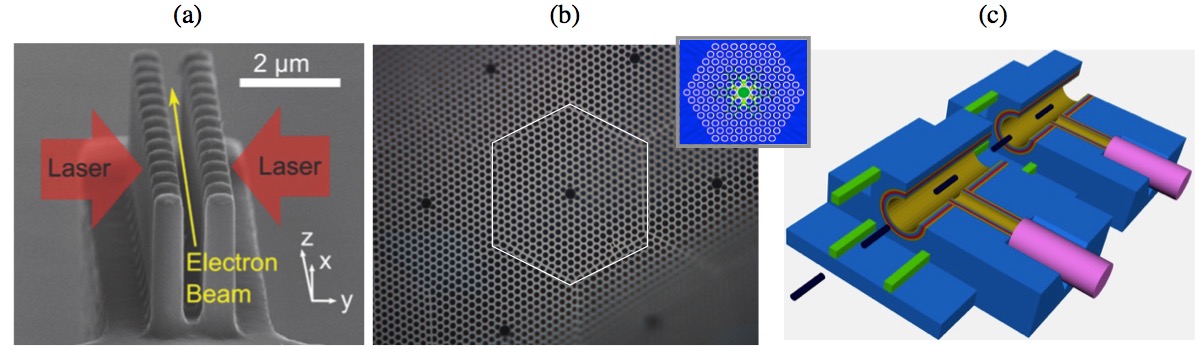}
\caption{Various DLA structures: (a) the planar-symmetric dual-pillar collonade geometry \cite{leedle:2018}; (b) a hexagonal hollow-core photonic crystal fiber geometry with TM$_{01}$ like mode (inset) \cite{noble:2011}; and (c) a proposed cylindrically symmetric Bragg geometry with cutaway revealing the interior \cite{dla:2011}.}
\label{structures}
\end{center}
 \end{figure}

The 2D (fiber) type geometry of Fig.~\ref{structures}(b) is similar conceptually to more conventional RF accelerators in that the confined accelerating mode is close to azimuthally symmetric, mimicking the transverse magnetic TM$_{01}$ mode of a conventional accelerating cavity.  However, the preferred fabrication technique (telecom fiber drawing) is less amenable to a fully on-chip approach. A lithographically produced variant of such an azimuthal structure, shown in Fig.~\ref{structures}(c), would allow confinement of a pure TM$_{01}$ accelerating mode, which is beneficial for stable beam transport over multi-meter distances. For such a structure, the radius of the vacuum channel is roughly $R = \lambda / 2$ whereas the bunch radius should satisfy $\sigma_r < 0.1 \lambda$, where $\lambda$ is the laser wavelength. This leads to two important aspects that should be emphasized. One is that due to constructive interference of a train of microbunches, the projection of the total wake on the fundamental mode is magnified. The other is that dielectric materials are virtually transparent over a large range of wavelengths. Consequently, whereas tens of thousands of modes are used for wake calculations in metallic RF cavities, only a few hundred modes contribute to long-range wake effects in a DLA. 



\subsubsection{Bunch Format}

In order to produce a net acceleration of charged particles using the DLA concept, the beam must be microbunched with a periodicity equal to the laser wavelength. Techniques for accomplishing this at relativistic beam energies have been previously demonstrated at wavelengths of 10 $\mu$m and 800 nm \cite{kimura:2001,sears:atto2008}. There is also the opportunity to bunch the beam already in the low energy injector~\cite{black:Atto:2019, schoenenberger:Atto:2019}, where special setups allow keeping the energy spread small~\cite{Niedermayer_Black_PhysRevApplied2020}. The luminosity requirement for a 30 TeV collider necessitates of order of 10$^{14}$ electrons per second at the interaction point. Current laser technology permits repetition rates as high as 100 MHz with pulse durations on the order of picoseconds. Consequently, each such laser pulse may contain an electron train of 100 to 1000 microbunches.  The bunch structure and the aperture of the acceleration structure determine the number of electrons to be contained in one microbunch. The effective radial field $E_\perp = (e n \sigma_r / 2 \epsilon_0 \gamma^3)$ generated by a pencil beam is suppressed by the relativistic Lorentz factor $\gamma$, but becomes very significant at low energies. For example, in the non-relativistic case $10^4$ electrons confined in a sphere of radius 0.1 $\mu$m  is of the order of 1 GV/m, which is comparable with the laser field. Consequently, one micro-bunch can not contain more than on the order of a few $10^4$ electrons.  In the example parameters of Table~\ref{parameters} we assume a microbunch charge of $3 \times 10^4$ electrons/positrons, which is also consistent with efficient multi-bunch operation.  

\subsubsection{Beam Break-Up (BBU)}

Beam breakup (BBU) instability was first observed in 1966 as the pulse length of the transmitted beam appeared to shorten, provided the beam current exceeds a threshold value at a given distance along the accelerator \cite{panofsky:BBU:1968}. This threshold is lower for longer distances. Its essentials were found to be transverse fields generated by the beam \cite{chao:1980}. During the years BBU attracted attention every time a new acceleration paradigm came to serious consideration: this was the case for NLC \cite{dehler:1998} and CLIC \cite{braun:2008}, where it has been suggested to damp and detune the structure (DDS) in order to suppress hybrid high order modes (HOM) that leads to BBU instability. Later when the energy recovery linac (ERL) was in focus BBU was investigated in this configuration \cite{hoffstaetter:2004} and further for superconducting RF gun \cite{volkov:2011}. Without exception, the acceleration structure is initially azimuthally symmetric and the hybrid modes are excited due to transverse offset or asymmetry of the beam or to the coupling of input or output arms. In the case of dielectric structures, with the exception of Bragg waveguide \cite{mizrahi:2004}, the acceleration modes are quasi-symmetric since the TM$_{01}$ mode is actually a hybrid mode with a transverse electric component, in addition to other possible hybrid modes. Consequently, for a linear collider, it may be desirable to use an azimuthally symmetric structure, such as the Bragg waveguide of Fig.~\ref{structures}(c), or to suppress dipole modes by careful structure design as discussed in Ref.~\cite{lin:2001} for the honeycomb photonic crystal geometry of Fig.~\ref{structures}(b). Tracking with wakefields, in order to assess the BBU in two-dimensional DLA structures was performed in Ref.~\cite{Egenolf_PRAB_2020}, where an extension of the DLAtrack6D tracking code \cite{niedermayer_beam_2017} allows the user to apply kicks of predetermined~\cite{CST} wake functions using a one-kick per DLA cell scheme.

\subsubsection{Emittance Requirements} 

As a figure of merit we keep in mind that to remove the beam sufficiently from the structure's wall we assumed that $\sigma_r \leq \lambda/10$. For a heuristic estimate of the geometric emittance we note that the transverse velocity should be smaller than the velocity required for an electron on axis at the input to hit the wall $r = R$ at the exit of one acceleration stage $z = L$. Therefore, the transverse emittance must be at least of the order $\epsilon_\text{rms} = R \sigma_\text{r} / L \simeq 0.5$ nm.  For the parameter tables of Section \ref{sec:parameters} we assume a normalized emittance of $\epsilon_N = \gamma \epsilon_\text{rms}$ = 0.1 nm at $\lambda$ = 2 $\mu$m. Since $\epsilon_N$ is preserved as $\gamma$ increases, we require that $\epsilon_N$ be matched at low energy ($\gamma \simeq 1$).


\subsubsection{Laser Power and Heat Dissipation}

Assuming an average loaded gradient of 1 GV/m, the active length of each arm of the collider is 1.5 km yielding approximately 0.1 MW/m of average laser beam power per accelerating channel, with 33 lasers per meter of active accelerator length and 1 kW of average power required per laser. While CW lasers exceed 50\% efficiency at slightly lower power levels, high-repetition rate lasers currently cannot deliver the necessary average power.  However, laser experts predict that modern fiber lasers will reach the requisite average power levels within 5--10 years \cite{dla:2011}.  Close concentration of such laser energy in a dielectric substrate raises concern about heat dissipation. Compared with metallic surfaces at RF frequencies, the absorption coefficients for dielectrics at optical wavelengths are relatively low. It is found in Ref.~\cite{karagodsky:2006} that, ignoring wake field effects, the heat dissipated by the fundamental mode in a Bragg waveguide is at least three orders of magnitude below the practical limit for thermal heat dissipation from planar surfaces (1500 W/cm$^2$). 


\subsubsection{Beam Transport and Focusing} 

For long-distance particle transport there is a serious need for a focusing system. Techniques for DLA have been proposed that utilize the laser field itself to produce a ponderomotive focusing force either by excitation of additional harmonic modes or by introducing drifts that alternate the laser field between accelerating and focusing phases to simultaneously provide acceleration as well as longitudinal and transverse confinement \cite{naranjo:2012,niedermayer:focusing:2018}. In simulation, such focusing techniques can adequately confine a particle beam to a narrow channel and overcome the resonant defocusing of the accelerating field. New structure designs and experiments are currently underway to test these approaches. Requiring that the confining force of a focusing lattice is stronger than the repelling force of the charged particles sets a limit on the total number of electrons in a bunch. The latter is determined by the momentum of the electrons and the energy density of the focusing system. In Ref.~\cite{hanuka:regimes:2018} the maximum charge is investigated that could be transported in four types of focusing lattices: Einzel lens, Solenoid, Electric or Magnetic quadrupole. While the Electric Quadrupole would facilitate the highest amount of charge, the applied voltage will be limited by the distance between two adjacent electrodes, such that breakdown is avoided. Therefore, it seems inevitable to split the bunch into a train of bunches in order to weaken the space-charge. This space-charge reduction comes at the expense of the maximum efficiency, which is at least 20\% lower than the single bunch configuration \cite{hanuka:multi:2018}. Another proposed scheme uses a radially polarized laser Bessel beam to transversely focus a counter-propagating particle beam \cite{schachter:kimura:2020}.
It is also possible to use Spatial Harmonic Focusing~\cite{naranjo_stable_2012} or Alternating Phase Focusing (APF)~\cite{niedermayer:focusing:2018} in order to waive external focusing requirements completely. However, a generalization to 3D of the originally proposed two-dimensional schemes is required. As discussed in~\cite{Niedermayer_PRL_2020}, the 3D scheme has advantages at low energy, where the confinement to the extremely small aperture can also be provided in the vertical axis. Recently, based on this scheme and using Silicon-on-Insulator (SOI) wafers, a completely scalable multi-stage accelerator on a chip could be designed~\cite{Niedermayer_PhysRevApplied_2021}. At high energy, the 3D APF scheme allows stronger focusing gradients, since only the square-sum of the two focusing constants vanishes with $\gamma^{-2}$~\cite{Niedermayer_PRL_2020}, such that in a counter-phase arrangement the two transverse planes can  exhibit focusing gradients that are not constrained by the beam energy. This allows designing structures of a high-damage-threshold single material, which allow shorter focusing periods and accept higher emittances than the preceeding 2D structures~\cite{Niedermayer_etal_this_issue}.

\subsection{Laser Coupling for Multi-stage Transport} 
\label{sec:coupling}


A major challenge of DLA is scaling up the interaction length between the driving laser and the electron beam, which is limited by both the beam dynamics and the laser delivery system. A promising solution is to use integrated optics platforms, built with precise nanofabrication, to provide controlled laser power delivery to the DLA, which would further eliminate many free-space optical components, which are bulky, expensive, and sensitive to alignment.  The laser control mechanisms may additionally be implemented on-chip, which will add to the compactness and robustness of the device and allow for precise implementation of laser-driven focusing schemes.

\begin{figure}
\begin{center}
 \includegraphics[height=.2\textheight]{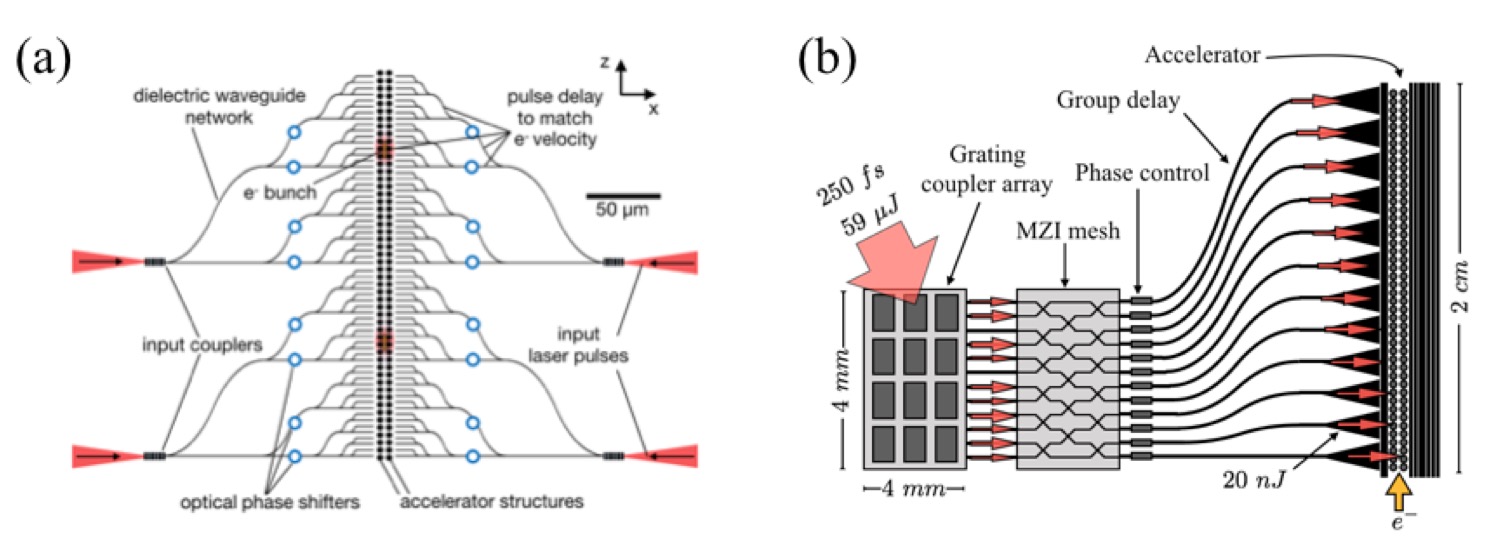}
\caption{(a) The DLA power distribution network concept proposed in \cite{hughes:chip:2018} using dielectric waveguides to split and delay a single input pulse to the accelerator structure, (b) A schematic of a proposed waveguide-fed DLA designed for long interaction length (not to scale).}
\label{treebranch}
\end{center}
 \end{figure}

A system for laser coupling to DLA was recently proposed in Ref.~\cite{hughes:chip:2018}, in which the laser beam is first coupled into a single dielectric waveguide on the chip and then split several times to spread over the accelerator structure.  Here, waveguide bends are designed to implement an on-chip pulse-front tilt, which delays the incident laser energy to arrive at the accelerator structure at the same time as the moving electron beam \cite{plettner_proposed_2006,wei:2017,cesar:pft:2018}.  While this work provides a way to achieving interaction lengths on the 100 $\mu$m to 1 mm scale, the power splitting approach has the disadvantage of concentrating the optical power at a single input facet.  For longer length structures, requiring more splits, the input facet becomes a bottleneck for damage and nonlinear effects, and future versions of DLA using integrated optical power delivery systems would ideally have ``one-to-many'' coupling mechanisms where a single laser beam is directly coupled into several waveguides, eliminating this bottleneck. One approach to this kind of coupling uses a large array of grating couplers on the surface of the chip, each supplying power to an individual waveguide, as shown in Fig.~\ref{treebranch}(b).  Theoretical studies of grating couplers, combined with inverse design optimization have shown that coupling efficiencies close to 100\% may be possible \cite{su:2018}.  With areas of several $\mu$m$^2$ having been demonstrated for grating couplers, several thousand may fit on a mm$^2$ area, which may easily be aligned with a free-space laser source.

To increase the robustness of the DLA coupling, an integrated mesh of Mach-Zehnder Interferometers (MZIs) could be fabricated onto the chip.  These MZIs have been experimentally demonstrated and act as tunable beamsplitters that may share power between waveguides as controlled by integrated optical phase shifters  \cite{miller:2015}.  Initial simulations suggest that for a 250 fs pulse, up to 10 MZIs can be accommodated, which would roughly correspond to being able to share power between 10 adjacent waveguides and should be sufficient for the purposes of DLA.  Phase control control may be accomplished by thermal or electro-optic phase shifters integrated on final waveguide sections.  Either the electron beam signal or the light scattered out of plane may be used as a diagnostic tool for sequentially optimizing the phase shifters.  These phase shifters may also be used to implement laser-driven focusing schemes, such as ponderomotive focusing \cite{naranjo:2012} or alternating phase focusing \cite{niedermayer:focusing:2018}, which have shown significant promise for DLA in recent simulation studies \cite{Niedermayer_PhysRevApplied_2021}.  Thus, integrated optical phase control gives a path forward for combined acceleration and focusing of the electron beam.

The group delay necessary for matching the arrival of each pulse to the moving electron bunch can be implemented by designing the fixed waveguide geometry, such as the bends described in \cite{hughes:chip:2018}, in combination with subwavelength gratings \cite{wang:2015} embedded in the waveguides.  With bend radii as low as 50$\mu$m, it is possible to get close to 100\% transmission through the bends \cite{hughes:chip:2018}.  Additional stages may be necessary for compensating dispersion encountered in the waveguides.  This will be especially important for longer structures.  Previous simulations \cite{hughes:chip:2018} showed that these effects will occur at around 1 cm waveguide lengths when using weakly-guided SiN waveguides.  An attractive option is to engineer this dispersion to avoid damage and nonlinearities, by sending in an initially chirped and broadened pulse, providing recompression closer to the accelerator.

The coupling from waveguide to several DLA periods may be accomplished with an inverse-taper on the waveguide.  Alternatively, the DLA structures may be etched directly into the waveguide, such as in a buried grating \cite{chang4}.  This method is currently being tested experimentally.  It was shown in Ref.~\cite{hughes:chip:2018} that a moderate amount of resonance may be beneficial for enhancing the electric fields in the accelerator gap and avoiding the damage and nonlinear constraints in the waveguides.  Thus, a quality factor of about 10 may be useful to design into the DLA structures either by inverse design using the adjoint method \cite{hughes:avm:2017} or by defining dielectric mirrors surrounding the DLA structures.  The entire structure may either be driven symmetrically on each side, or, alternatively, a dielectric mirror may be used to reflect the incoming light from one side of the device.

\subsection{Laser Requirements}  
\label{sec:lasers}
{\hskip 0.13in}
The drive laser requirements for a DLA based accelerator reflect the power and efficiency requirements for future real-world applications as well as the unusual pulse format of the electron beam: namely, pulse energies in the range of 1 to 10 $\mu$J, with $\leq$ 1 ps pulse duration, 10 to 100 MHz repetition rates, and high (> 30\%) wall-plug efficiencies. In addition, the optical phase of the base carrier wave needs to be locked to the phase of the accelerating electron beam.  The nominal laser type will probably be a fiber laser because of its efficiency and robust, low maintenance operation.  Fiber lasers with 1 $\mu$m wavelength and hundreds of Watts of average power have already been demonstrated to be capable of meeting most of these parameter requirements, and higher power (>1 kW) mode-locked systems at longer wavelengths (e.g. 2 $\mu$m Thulium-doped lasers) are now commercially available. Consequently, the current state of the art in laser systems is not far from what will eventually be required for large-scale accelerators based upon DLA.

\begin{figure}
\begin{center}
 \includegraphics[height=.25\textheight]{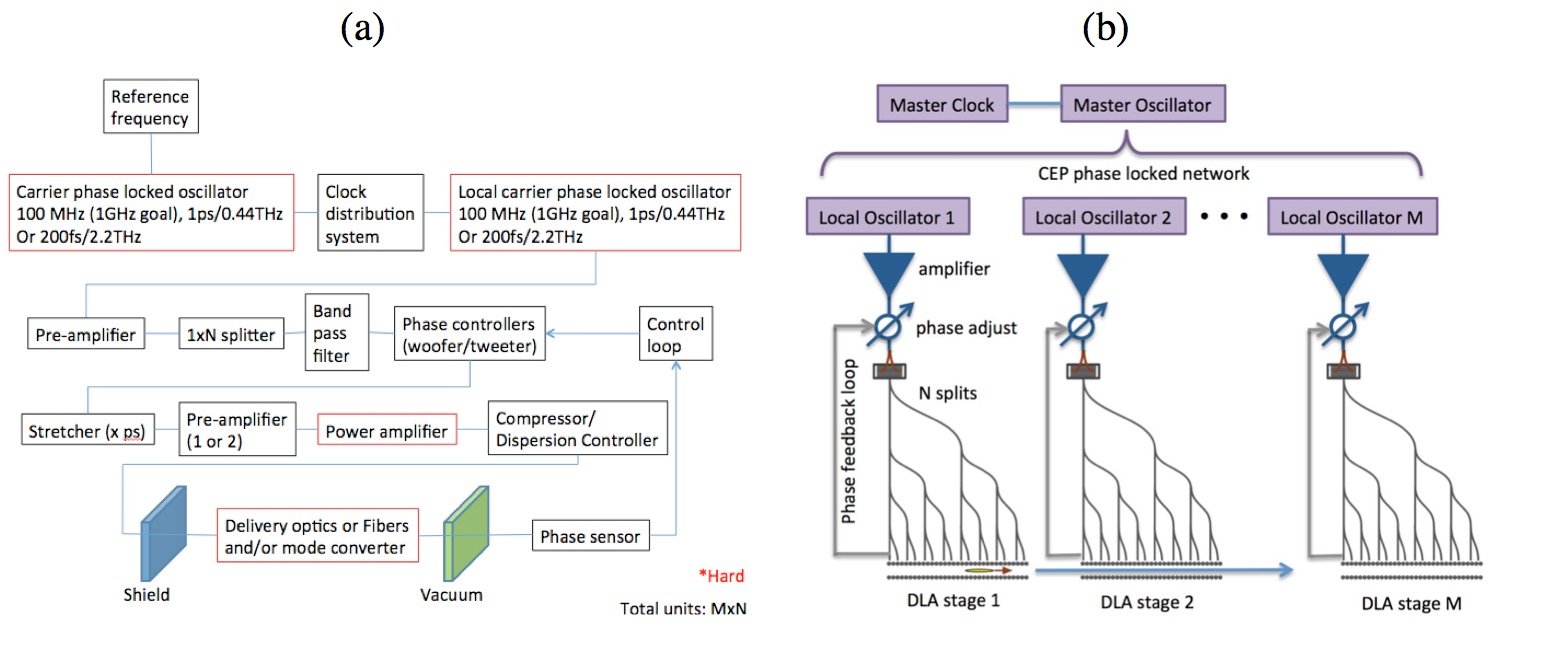}
\caption{Conceptual laser system baseline design for a multi-stage DLA accelerator with $M$ local oscillators, showing (a) block diagram as outlined in Ref.~\cite{dla:2011} and (b) a simplified component-level schematic.}
\label{laser_schematic}
\end{center}
 \end{figure}

A baseline conceptual design for a laser system for a multi-stage DLA is shown in Fig.~\ref{laser_schematic}, adapted from \cite{dla:2011}. The design is modular to enable scaling to higher energies with more stages, with timing across a long accelerator as one of the significant technical challenges. The baseline design begins by producing a carrier envelope phase (CEP)-locked oscillator with its repetition rate matched to a stable RF reference frequency source in the range of 100 MHz to 1 GHz, with 1 GHz being the target. This oscillator will serve as the clock for the accelerator. The global oscillator or clock will be distributed via optical fiber to local oscillators, which are phase-locked to the global oscillator. Each structure will require a phase control loop to allow for acceleration through successive structures. Both fast and slow control of the phase will be necessary. By monitoring the energy linewidth as well as the timing of the electron bunches, successful acceleration through the structures may be confirmed.

The baseline conceptual design looks to be a manageable system, with the toughest challenges coming from the requirements for the oscillators, the power amplifier, and the delivery optics. In addition, it will be necessary to repeat the local system multiple times, with each local system phase-locked to the global oscillator. We emphasize that the conceptual illustration of Fig.~\ref{laser_schematic} is our envisioning of a fully scalable laser drive network for a multi-stage accelerator. The main areas of development needed for the laser technology development are timing accuracy and distribution (combined with phase sensing and feedback at the point light is coupled to the electron beam), and beam transport and coupling of the laser to the accelerator structure. Consequently, the key areas of technology development required are: (1) development and demonstration of scalable techniques for sub-cycle phase-locking of multiple fiber lasers, and (2) numerical design, fabrication, and benchtop testing of ultracompact delivery optics. 

As the acceleration process of DLA is linear with the electric field, the optical phase must be well controlled. Poor synchronization would result in either a decrease of efficiency or an electron energy spreading or even defocusing. Frequency comb technologies can detect and control both the repetition rate of the delivered pulses and the carrier to envelop phase (CEP). Frequency comb techniques can be used to precisely measure and fine-tune repetition rate and CEP of delivered pulses. Because the electrons in a dielectric laser accelerator will be optically compressed to form microbunches less than one optical cycle in duration and separated by a single laser wavelength, this requires sub-cycle stabilization of the absolute frequency and relative CEP of each pulse. 

\subsection{Beam Delivery System }

Beam delivery for a DLA based linac could take advantage of the same techniques developed for modern colliders, including conventional collimation systems and magnetic optics at the final focus. The primary purpose of the collimation system is to protect the detector from background due to the particle spray from beam halo that intercept the apertures of the accelerator. However, due to the extremely low-charge, low-emittance bunches that would be produced and the narrow beam apertures, a DLA electron beam would already be inherently collimated to less than a micron in size. Consequently the need for an additional long collimation region prior to the IP may be significantly mitigated. In the parameters of Table~\ref{parameters} we assume a collimation length of 1.92 km prior to the final focus section. This section could simultaneously employ dispersive elements to wash out the microbunches so that the collided beams at the IP correspond to ps duration bunches with the combined charge of the bunch train. This allows for a luminosity enhancement of approximately 10 due to the pinching effect at the IP from the beam-beam interaction. Since the gross transverse dimensions of a DLA accelerator would be extremely small (of order 1 mm), particle spray from halo could also potentially be intercepted or angularly dispersed by shielding around the linac itself. The drift length $L^*$ to the detector is difficult to estimate absent details of the detector design. In Table~\ref{parameters} we assumed a drift $L^*$ = 5 m, similar to ILC values.

\section{Integrated system}

\subsection{Design Tolerances}
The Dielectric Laser Accelerator Workshop held at SLAC National Accelerator Laboratory in 2011 examined required tolerances for a DLA based collider \cite{dla:2011}. Since the emittance must be preserved through several kilometers of acceleration, misalignments must be small enough that they do not result in significant emittance growth. Conventional magnetic focusing would require tolerances of order 1 $\mu$m in quadrupole magnet positioning, 100 nm in the accelerator structure alignment, and quadrupole jitter of less than 0.1 nm. This was based on requiring a maximum centroid motion of 10\% of the beam size from magnetic center vibration, assuming 1000 quads and a normalized transverse emittance of 0.1 nm. However, proposed electromagnetic focusing schemes which are now being incorporated into structure designs and experiments, such as alternating phase focusing and nonresonant harmonic focusing \cite{naranjo_stable_2012,niedermayer:focusing:2018}, can be built into the DLA structure design with nanometric precision that should well exceed such tolerances. A preliminary study of beam breakup instability (BBU) using a simple two-particle model found that a 30 nm average misalignment resulted in a transverse normalized emittance growth of 2.2 nm from a cold beam over 500 GeV of acceleration in 1 kilometer. A scan of emittance growth vs. bunch charge was conducted, and it was found that accelerating sufficient charge with tolerable beam degradation for high-energy physics applications requires about 50 nm alignment. Beam stability may be improved by using a shorter focusing period or by use of Balakin-Novokhatsky-Smirnov (BNS) damping. While achieving such tolerances over several kilometers is challenging, the high repetition rate of a DLA collider provides information at MHz frequencies, which can be used for feedback stabilization. Stabilization of optical components to better than 1 nm Hz$^{-1/2}$ has already been demonstrated over similar length scales at the LIGO facility \cite{ligo:stability}. Furthermore, since the acceleration process of DLA is linear with the electric field, the optical phase must be well controlled. Frequency comb technologies can detect and control both the repetition rate of the delivered pulses and the carrier to envelope phase (CEP). The technology used to generate frequency combs in ultra-high finesse Fabry Perot cavities is able to control phase noise in the range of 0.01 Hz to 100 kHz. Further stabilization will necessitate control systems operating above 100 kHz and requires important efforts in feedback loop electronics and ultrafast low-noise detectors. 

\subsection{Instrumentation}

A future DLA-based linear collider will require the development suitable diagnostics and beam manipulation techniques, including compatible small-footprint deflectors, focusing elements, and beam position monitors (BPMs), concepts for which have been proposed \cite{plettner:2008,plettner:2011,soong:2012,naranjo:2012,soong:2012b}.  A dielectric laser-driven element that produces transverse deflection forces in both transverse dimensions as well as a longitudinal accelerating force was proposed by Plettner and Byer \cite{plettner:2008}. The concept is similar to the planar symmetric grating accelerator but with the gratings tilted at an angle relative to the particle beam axis. By exciting a superposition of TE and TM modes in such a deflector with a tilt angle of 45$^\circ$, it has been recently shown that a pure deflection mode can be excited (all other force components cancel) which could be used to make nanoscale orbit corrections, to make a laser driven undulator, or to steer the electron beam between successive chips or wafers \cite{englandAAC:2018}. 

A proof-of-principle demonstration of the BPM concept was recently conducted  \cite{soong:2014}.  The concept uses a dual-grating with a tapered grating period to produce a linear variation in operating wavelength along the dimension transverse to the electron beam axis.  Light emitted by wakefield excitation of the electron beam (via the inverse of the acceleration process) has a different center wavelength depending on transverse position of the electrons, permitting a measurement of beam position from the power spectrum of emitted light.  Similarly the spectral width of the emitted radiation is a measure of the beam size, permitting a simultaneous determination of beam position and transverse size.  The direction of the variation is chosen to correspond to either horizontal or vertical offset of the particle beam. When a particle beam traverses this BPM structure, it generates radiation at a wavelength corresponding to the grating period at the beam location. Combined with a DLA-generated sub-micron sized electron beam and a typical 0.1 nm resolution spectrograph, it is estimated that this technique could be employed to resolve beam position in a DLA with 0.75 nm precision. 

\subsection{Simulation}
\label{sec:simulation}
{\hskip 0.13in}
The DLA scheme is analogous in many respects to a frequency-scaled version of a conventional accelerator.  Consequently, modelling many aspects of the particle dynamics and transport in a DLA collider can rely largely upon well established accelerator codes and computational methods.  This reduces the need for new code developments or large-scale computing infrastructure to a subset of DLA simulation tasks.  For simulation purposes, a linear collider can be broken into three regimes which require different kinds of modeling:  (1) Injector (source, emittance preparation, and acceleration up to 1 GeV); (2) Accelerator (1 GeV to 30 TeV) and (3) beam delivery, including final focus, crab crossing, IP design, and beamstrahlung.  The proposed injection scheme outlined in Section \ref{sec:parameters} is based on superconducting RF technology, and thus is amenable to the same computational methods used to model such systems and will therefore not be addressed here. The main accelerator needs longitudinally coupled structures (traveling wave structures) for energy efficiency and mode stability reasons as described in Section \ref{sec:structures}. The main accelerator can then be split hierarchically into separately addressed sections corresponding to their respective length scales: one optical period (2 $\mu$m), power coupling cell (10 cm), focusing cell (20 m at 3 TeV), one betatron period (100 m at 3 TeV), bunch compression stage (1 km), full linac (4.2 km each side at 3 TeV).

An individual DLA cell (one optical period in length) can be straightforwardly simulated by any of a variety of finite difference time domain (FDTD), finite difference frequency domain (FDFD), and finite element method (FEM) codes combined with various optimization techniques \cite{shin:2013,egenolf:2017,hughes:avm:2017}. The resulting single-cell Fourier coefficients, combined with the corresponding phase and group velocities, can then be combined with 6D tracking codes to obtain long-distance phase space evolutions by symplectic one-kick-per-cell tracking \cite{niedermayer:2017}.  Advanced codes and large-scale computing become necessary for the integration of particle trajectories through longer structure segments.  Compared to conventional accelerators, DLAs have much smaller feature sizes, which must be resolved in simulations, leading to much more demanding computations.  Particle-in-cell (PIC) simulations can be used to model full particle and field dynamics up to a few thousand or tens of thousands of structure periods but becomes computationally prohibitive at longer length scales.  At optical-scale frequencies, a 10 cm interaction distance would be near the limit of what is possible with a modern supercomputer.  A process of using simplified models to construct larger building blocks can be applied to successive levels of the design, using transfer maps to represent larger-scale components such as an entire power coupling cell or focusing cell. 

Moreover, wake functions and beam coupling impedances must be included in the various structures, which can be done both in full 3D~\cite{CST,VSim} or by simplified 2D models in the frequency domain~\cite{Niedermayer_PRAB_2015}. The most crucial issues here are to find the beam loading and beam break up limits using the longitudinal and transverse wakes~\cite{Egenolf_PRAB_2020}. For the extremely short sub-optical bunches employed in a DLA scenario (with bunch length small compared to transverse size) the theory of beam instabilities might require extension by nonlinear parts of the transverse wakes and transverse position dependent parts of the longitudinal wakes. Such theories should be validated by extensive PIC simulations. Moreover, the consequences of slight steering errors in the 10-nanometer range, leading to severe average beam power deposition in the structures needs to be studied. Already available radiation damage codes such as FLUKA~\cite{FLuka} can be applied here.  While much of the above simulation work can be accomplished by use of existing codes and/or well-established computational techniques, the beam dynamics in the final focus of a DLA collider needs to be completely redeveloped for DLA-type beams, unless, as assumed in Section \ref{sec:parameters}, the microbunching is washed out prior to interaction at the IP.  Direct collision of trains of extremely short, low intensity bunches with high repetition rate would behave very differently than conventional bunch crossings when it comes to beamstrahlung or crab crossing.  Fast beam-beam interaction codes are available, but have yet to be adapted to attosecond bunches.

\section{Partners, Resources, and Current Activity Level.}
\label{sec:partners}

Progress towards an energy scalable architecture based upon laser acceleration in dielectric materials requires an R\&D focus on fabrication and structure evaluation to optimize existing and proposed concepts, and development of low-charge high-repetition-rate particle sources that can be used to demonstrate performance over many stages of acceleration. To tackle these challenges, a concerted effort is required that leverages industrial fabrication capabilities and that draws upon world-class expertise in multiple areas. A number of university, national laboratory, and industrial institutions and collaborations are now actively conducting research in this area, including the multi-institutional Accelerator on a Chip International Program (ACHIP), which includes 6 universities, 1 company, and 3 national laboratories, as well as Los Alamos National Laboratory, University of Tokyo, Tel-Aviv University, The Technion, and University of Liverpool.  The DLA effort could be made technology-limited through appropriate leveraging of the semiconductor and laser R\&D industries and appropriate growth of the DLA research community.  Current collaborative efforts in the U.S. and Europe are aimed at developing a first R\&D demonstration system incorporating multiple stages of acceleration, efficient guided wave systems, and high repetition rate solid state laser systems.

Testing of speed-of-light prototype devices and initial staging experiments will require suitable test facilities equipped with relativistic beams. Existing conventional RF facilities are suitable for near-term tests over the next few years.  To provide an estimate of projected facilities costs, the current ACHIP program includes roughly \$1.3M/year of combined in-kind support from three national laboratories (SLAC, DESY, and PSI) in the form of access to personnel, resources, and beam time. However, for demonstrating many-staged DLA accelerators, ultra-low emittance particle sources need to be developed and combined with DLA devices to make compatible injectors. Ongoing development of DLA prototype integrated systems will provide a pathway for scaling of this technology to MeV, GeV, and then TeV energies and to beam brightnesses of interest both for high energy physics and for a host of other applications, as discussed in Ref.~\cite{england:review:2016}.  

\section{Conclusion}
As an advanced accelerator concept, the DLA approach offers some unique advantages: the acceleration mechanism is inherently linear and occurs in a vacuum region in a static structure.  In addition to the stability benefits this affords, it also means that the acceleration effect is inherently dependent on the phase of the laser field, which makes it possible to dynamically fine-tune accelerator performance by manipulation of the incident laser phase profile.  Due to its low-charge high repetition rate bunch format, the projected beamsstrahlung energy loss for a multi-TeV scenario is in the few percent range, as opposed to tens of percents for conventional RF accelerators.  Gradients on the GV/m scale have already been demonstrated, and wall plug efficiencies comparable or superior to conventional approaches appear quite feasible.  Furthermore, the primary supporting technologies (solid state lasers and nanofabrication) are already at or near the capabilities required for a full-scale accelerator based on this approach. These advantages motivate supporting DLA research as a competitive higher gradient alternative to more conventional RF accelerators.  While synergistic applications (i.e., those needing ultra-compact accelerators) provide an opportunity for cost-sharing the technology development, overall the DLA technology has had significantly lower funding levels than other advanced accelerator technologies and essentially no funding dedicated to closing technology gaps unique for multi-TeV linear colliders. This is partly due to the perception that DLA is not a credible technology path for a 30 MW linear collider beam due to its small, micron-sized aperture, even when dividing the main beam power into multiple, parallel DLA lines each with lower individual power. This perception needs to be addressed as early in the DLA technology roadmap as possible. The key technology demonstrations needed for a DLA linear collider development in the near term include: (1) demonstration of low-cost high-efficiency dielectric structures with sufficient thermal conductivity and controllable wakefield effects; (2) demonstration of focusing schemes with sufficiently high gradient to minimize beam interception without diluting the beam quality; and (3) demonstration of low-cost, high-efficiency, and high-power drive lasers that have sub-cycle phase and timing control. Successful demonstration of these technologies will show that the DLA technology is a credible alternative for a multi-TeV linear collider design and will provide the motivation for a multi-stage linear collider prototype within 20 years.

\acknowledgments
This work was supported in part by the Gordon and Betty Moore Foundation (GBMF4744), the U.S. Department of Energy (DE-AC02-76SF00515), the German Ministry of Education and Research (Grant No. 05K19RDE), and the state of Hessen via LOEWE Exploration,  the Israel Science
Foundation (ISF) and STI Optronics, Inc. Internal
Research and Development.


\bibliographystyle{unsrt}
\bibliography{england}

\begin{thebibliography}{10}

\bibitem{wootton_demonstration_2016}
Kent~P. Wootton, Ziran Wu, Benjamin~M. Cowan, Adi Hanuka, Igor~V. Makasyuk,
  Edgar~A. Peralta, Ken Soong, Robert~L. Byer, and R.~J. England.
\newblock Demonstration of acceleration of relativistic electrons at a
  dielectric microstructure using femtosecond laser pulses.
\newblock {\em Opt. Lett.}, 41(12):2696, June 2016.

\bibitem{cesar_nonlinear_2018}
D.~Cesar, S.~Custodio, J.~Maxson, P.~Musumeci, X.~Shen, E.~Threlkeld, R.~J.
  England, A.~Hanuka, I.~V. Makasyuk, E.~A. Peralta, K.~P. Wootton, and Z.~Wu.
\newblock High-field nonlinear optical response and phase control in a
  dielectric laser accelerator.
\newblock {\em Nature Comm. Phys.}, 1(4):1--7, August 2018.

\bibitem{peralta:2013}
E.~A. Peralta, K.~Soong, R.~J. England, E.~R. Colby, Z.~Wu, B.~Montazeri,
  C.~McGuinness, J.~McNeur, K.~J. Leedle, D.~Walz, E.~B. Sozer, B.~Cowan,
  B.~Schwartz, G.~Travish, and R.~L. Byer.
\newblock Demonstration of electron acceleration in a laser-driven dielectric
  microstructure.
\newblock {\em Nature}, 503:91--94, 2013.

\bibitem{cesar:pft:2018}
D.~Cesar, J.~Maxson, X.~Shen, K.~P. Wootton, S.~Tan, R.~J. England, and
  P.~Musumeci.
\newblock {Enhanced energy gain in a dielectric laser accelerator using a
  tilted pulse front laser}.
\newblock {\em Optics Express}, 26:29216, 2018.

\bibitem{breuer_laser-based_2013}
John Breuer and Peter Hommelhoff.
\newblock Laser-{Based} {Acceleration} of {Nonrelativistic} {Electrons} at a
  {Dielectric} {Structure}.
\newblock {\em Phys. Rev. Lett.}, 111(13):134803, September 2013.

\bibitem{leedle_dielectric_2015}
Kenneth~J. Leedle, Andrew Ceballos, Huiyang Deng, Olav Solgaard, R.~F. Pease,
  Robert~L. Byer, and James~S. Harris.
\newblock Dielectric laser acceleration of sub-100 {keV} electrons with silicon
  dual-pillar grating structures.
\newblock {\em Opt. Lett.}, 40(18):4344, September 2015.

\bibitem{hughes:chip:2018}
Tyler~W. Hughes et~al.
\newblock On-{Chip} {Laser}-{Power} {Delivery} {System} for {Dielectric}
  {Laser} {Accelerators}.
\newblock {\em Phys. Rev. Applied}, 9:054017, May 2018.

\bibitem{mcneur_elements_2018}
Joshua McNeur, Martin Koz\'{a}k, Norbert Sch\"{o}nenberger, Kenneth~J. Leedle,
  Huiyang Deng, Andrew Ceballos, Heinar Hoogland, Axel Ruehl, Ingmar Hartl,
  Ronald Holzwarth, Olav Solgaard, James~S. Harris, Robert~L. Byer, and Peter
  Hommelhoff.
\newblock Elements of a dielectric laser accelerator.
\newblock {\em Optica}, 5(6):687--690, April 2018.

\bibitem{colby:2011}
E.~R. Colby, R.~J. England, and R.~J. Noble.
\newblock A laser-driven linear collider: Sample machine parameters and
  configuration.
\newblock In {\em 2011 Particle Accelerator Conference Proceedings}, page 262,
  New York, NY, 2011. PAC'11 OC/IEEE.

\bibitem{siemann:2004}
R.~Siemann.
\newblock Energy efficiency of laser driven structure based accelerators.
\newblock {\em Phys. Rev. ST Accel. Beams}, 7:061303, 2004.

\bibitem{dla:2011}
P.~Bermel, R.~L. Byer, E.~R. Colby, B.~M. Cowan, J.~Dawson, R.~J. England,
  R.~J. Noble, M.~Qi, and R.~B. Yoder.
\newblock Summary of the 2011 dielectric laser accelerator workshop.
\newblock {\em Nucl. Instr. Meth. Phys. Res. A}, 734A:51--59, 2014.

\bibitem{snowmass:2013}
M.~Battaglia et~al.
\newblock Energy frontier lepton and photon colliders, section 31.
\newblock In {\em Proc. of Community Summer Study (CSS/Snowmass)}, pages
  20--21, 2013.

\bibitem{england:rmp2014}
R.~J. England, R.~J. Noble, et~al.
\newblock Dielectric laser accelerators.
\newblock {\em Rev. Mod. Phys.}, 86:1337, 2014.

\bibitem{beambeam:2021}
R.~J. England and L.~Schachter.
\newblock {Beam-beam interaction in a dielectric laser accelerator
  electron-positron collider}.
\newblock {\em Phys. Rev. Accel. Beams}, 24:121302, December 2021.

\bibitem{anar:2017}
B.~Cros and P.~Muggli.
\newblock {\em {Towards a Proposal for an Advanced Linear Collider: Report on
  the Advanced and Novel Accelerators for High Energy Physics Roadmap Workshop
  (ANAR 2017)}}.
\newblock CERN, Geneva, Switzerland, September 2017.

\bibitem{hulme:2014}
J.~C. Hulme et~al.
\newblock Fully integrated hybrid silicon two dimensional beam scanner.
\newblock {\em Opt. Exp.}, 23(5):5861--5874, 2014.

\bibitem{xiang:2016}
C.~Xiang, M.~A. Tran, T.~Komlenovic, J.~Hulme, M.~Davenport, D.~Baney,
  B.~Szafraniec, and J.~E. Bowers.
\newblock Integrated chip-scale {{Si3N4}} wavemeter with narrow free spectral
  range and high stability.
\newblock {\em Opt. Lett.}, 41(14):3309--3312, 2016.

\bibitem{ligo:stability}
F.~Matichard et~al.
\newblock {Advanced LIGO two-stage twelve-axis vibration isolation and
  positioning platform. Part 1: Design and production overview}.
\newblock {\em Precis. Eng.}, 40:273--286, 2015.

\bibitem{breuer:2013}
J.~Breuer and P.~Hommelhoff.
\newblock Laser-based acceleration of non-relativistic electrons at a
  dielectric structure.
\newblock {\em Phys. Rev. Lett.}, 111:134803, 2013.

\bibitem{leedle:2015}
K.~J. Leedle, R.~F. Pease, R.~L. Byer, and J.~S. Harris.
\newblock {Laser Acceleration and Deflection of 96.3 keV Electrons with a
  Silicon Dielectric Structure}.
\newblock {\em Optica}, 2:158--161, 2015.

\bibitem{england:review:2016}
R.~J. England.
\newblock Review of laser-driven photonic structure-based particle
  acceleration.
\newblock {\em IEEE J. Sel. Top. Quantum Electron.}, 22(2):4401007, March 2016.

\bibitem{P5:2014}
Steve Ritz et~al.
\newblock {Building for Discovery: Strategic Plan for U. S. Particle Physics in
  the Global Context}.
\newblock Technical report, DOE/HEP Particle Physics Project Prioritization
  Panel (P5) Report, May 2014.

\bibitem{rast:2016}
A.~W. Chao and W.~Chou.
\newblock {\em {Reviews of Accelerator Science and Technology}}, volume~9.
\newblock World Scientific Publishing Company, Geneva, Switzerland, January
  2016.

\bibitem{king:2000}
Bruce~J. King.
\newblock {Prospects for Colliders and Collider Physics to the 1 PeV Energy
  Scale}.
\newblock Technical Report {BNL-67410, CAP-282-Muon-00C}, Brookhaven National
  Laboratory, Upton, NY, April 2000.

\bibitem{schachter:kimura:2017}
L.~Schachter and W.~D. Kimura.
\newblock Quasi-monoenergetic ultrashort microbunch electron source.
\newblock {\em Nucl. Instrum. Methods. Phys. Res. Sect. A}, 875:80--86, 2017.

\bibitem{arnold:2011}
A.~Arnold and J.~Teichert.
\newblock Overview on superconducting photoinjectors.
\newblock {\em Phys. Rev. ST Accel. Beams}, 14:024801, 2011.

\bibitem{shimoda:1962}
K.~Shimoda.
\newblock Proposal for an electron accelerator using an optical maser.
\newblock {\em Appl. Opt.}, 1:33--35, 1962.

\bibitem{takeda:1968}
Y.~Takeda and I.~Matsui.
\newblock Laser linac with grating.
\newblock {\em Nucl. Instrum. Methods}, 62:306--310, 1968.

\bibitem{palmer:1980}
R.~B. Palmer.
\newblock A laser-driven grating linac.
\newblock {\em Part. Accel.}, 11:81, 1980.

\bibitem{palmer:1982}
R.~B. Palmer.
\newblock Near field accelerators.
\newblock In {\em Laser Acceleration of Particles, Los Alamos 1982}, page 179.
  AIP Conference Proceedings 91, AIP New York, 1982.

\bibitem{kroll:1985}
N.~M. Kroll.
\newblock General features of accelerating modes in open structures.
\newblock In {\em Laser Acceleration of Particles, Malibu 1985}, page 253. AIP
  Conference Proceedings 130, AIP New York, 1985.

\bibitem{pickup:1985}
M.~Pickup.
\newblock A grating linac at microwave frequencies.
\newblock In {\em Laser Acceleration of Particles, Malibu 1985}, page 281. AIP
  Conference Proceedings 130, AIP New York, 1985.

\bibitem{leap:2005}
T.~Plettner, R.~L. Byer, E.~R. Colby, C.~M.~S. Sears, J.~Spencer, and R.~H.
  Siemann.
\newblock Visible-laser acceleration of relativistic electrons in a
  semi-infinite vacuum.
\newblock {\em Phys. Rev. Lett.}, 95:134801, 2005.

\bibitem{sears:2008}
C.~M.~S. Sears, E.~Colby, R.~J. England., R.~Ischebeck, C.~McGuinness,
  J.~Nelson, R.~Noble, R.~H. Siemann, J.~Spencer, D.~Walz, T.~Plettner, and
  R.~L. Byer.
\newblock Phase stable net acceleration of electrons from a two-stage optical
  accelerator.
\newblock {\em Phys. Rev. ST Accel. Beams}, 11:101301, 2008.

\bibitem{rosing:1990}
M.~Rosing and W.~Gai.
\newblock {\em Phys. Rev. D}, 42:1829, 1990.

\bibitem{lin:2001}
X.~E. Lin.
\newblock Photonic band gap fiber accelerator.
\newblock {\em Phys. Rev. ST Accel. Beams}, 4:051301, 2001.

\bibitem{cowan:2003}
B.~M. Cowan.
\newblock {\em Phys. Rev. ST Accel. Beams}, 6:101301, 2003.

\bibitem{mizrahi:2004}
A.~Mizrahi and L.~Schachter.
\newblock {Optical Bragg Accelerators}.
\newblock {\em Phys. Rev. E}, 70:016505, 2004.

\bibitem{schachter:2004}
L.~Schachter.
\newblock {\em Phys. Rev. E}, 70:016504, 2004.

\bibitem{naranjo:2012}
B.~Naranjo, A.~Valloni, S.~Putterman, and J.~B. Rosenzweig.
\newblock {\em Phys. Rev. Lett.}, 109:164803, 2012.

\bibitem{scheuer:2014}
D.~Bar-Lev and J.~Scheuer.
\newblock Plasmonic metasurface for efficient ultrashort pulse laser-driven
  particle acceleration.
\newblock {\em Phys. Rev. ST Accel. Beams}, 17:121302, 2014.

\bibitem{mcneur:2012}
J.~McNeur, N.~Carranza, G.~Travish, H.~Yin, and R.~B. Yoder.
\newblock In R.~Zgadzaj, editor, {\em Advanced Accelerator Concepts, 15th
  Workshop, 2012}, AIP Conf.\ Proc.\ vol.\ 1507, pages 464--469, New York,
  2012. American Institute of Physics.

\bibitem{hanuka:single:2018}
A.~Hanuka and L.~Schachter.
\newblock {Optimized Operation of Dielectric Laser Accelerators: Single Bunch}.
\newblock {\em Phys. Rev. Accel. Beams}, 21:54001, 2018.

\bibitem{hanuka:regimes:2018}
A.~Hanuka and L.~Schachter.
\newblock {Operation regimes of a dielectric laser accelerator}.
\newblock {\em Nucl. Instr. Meth. Phys. Res. Sect. A}, 888:147--152, 2017.

\bibitem{na:2005}
Y.~C.~N. Na, R.~Siemann, and R.~L. Byer.
\newblock {\em Phys. Rev. ST Accel. Beams}, 8:031301, 2005.

\bibitem{moulton:2009}
P.~F. Moulton, G.~A. Rines, E.~Slobodtchikov, K.~F. Wall, G.~Frith, B.~Sampson,
  and A.~Carter.
\newblock Tm-doped fiber lasers: fundamentals and power scaling.
\newblock {\em IEEE J. Sel. Top. Quantum Electron.}, 15:85, 2009.

\bibitem{wu:2014}
Z.~Wu, R.~J. England, C-K. Ng, B.~Cowan, C.~McGuinness, C.~Lee, M.~Qi, and
  S.~Tantawi.
\newblock Coupling power into accelerating mode of a three-dimensional silicon
  woodpile photonics band-gap waveguide.
\newblock {\em Phys. Rev. ST Accel. Beams}, 17:081301, 2014.

\bibitem{leedle:2018}
K.~J. Leedle et~al.
\newblock Phase-dependent laser acceleration of electrons with symmetrically
  driven silicon dual pillar gratings.
\newblock {\em Opt. Lett.}, 43(9):2181--2184, 2018.

\bibitem{noble:2011}
R.~J. Noble, J.~E. Spencer, and B.~T. Kuhlmey.
\newblock Hollow-core photonic band gap fibers for particle acceleration.
\newblock {\em Phys. Rev. ST Accel. Beams}, 14:121303, 2011.

\bibitem{kimura:2001}
W.~Kimura, A.~vanSteenbergen, M.~Babzien, I.~Ben-Zvi, L.~P. Campbell, D.~B.
  Cline, C.~E. Dilley, J.~C. Gallardo, S.~C. Gottschalk, P.~He, K.~P. Kusche,
  Y.~Liu, R.~H. Pantell, I.~V. Pogorelsky, D.~C. Quimby, J.~Skaritka, L.~C.
  Steinhauer, and V.~Yakimenko.
\newblock {\em Phys. Rev. Lett.}, 86:4041, 2001.

\bibitem{sears:atto2008}
C.~M.~S. Sears, E.~Colby, R.~Ischebeck, C.~McGuinness, J.~Nelson, R.~Noble,
  R.~H. Siemann, J.~Spencer, D.~Walz, T.~Plettner, and R.~L. Byer.
\newblock Production and characterization of attosecond electron bunch trains.
\newblock {\em Phys. Rev. ST Accel. Beams}, 11:061301, 2008.

\bibitem{black:Atto:2019}
D.~S. Black et~al.
\newblock {Net Acceleration and Direct Measurement of Attosecond Electron
  Pulses in a Silicon Dielectric Laser Accelerator}.
\newblock {\em Phys. Rev. Lett.}, 123:264802, 2019.

\bibitem{schoenenberger:Atto:2019}
N.~Schoenenberger, Anna Mittelbach, Peyman Yousefi, Joshua McNeur, Uwe
  Niedermayer, and Peter Hommelhoff.
\newblock {Generation and Characterization of Attosecond Microbunched Electron
  Pulse Trains via Dielectric Laser Acceleration}.
\newblock {\em Phys. Rev. Lett.}, 123:264803, 2019.

\bibitem{Niedermayer_Black_PhysRevApplied2020}
U.~Niedermauyer, D.~S. Black, K.~J. Leedle, Y.~Miao, R.~L. Byer, and
  O.~Solgaard.
\newblock {Low-Energy-Spread Attosecond Bunching and Coherent Electron
  Acceleration in Dielectric Nanostructures}.
\newblock {\em Phys. Rev. Appl.}, 15:L021002, February 2021.

\bibitem{panofsky:BBU:1968}
W.~K.~H. Panofsky and M.~Bander.
\newblock {Asymptotic theory of beam break-up in linear accelerators}.
\newblock {\em Rev. Sci. Intrum.}, 39(2):206--212, 1968.

\bibitem{chao:1980}
A.~W. Chao, B.~Richter, and C.~Y. Yao.
\newblock {Beam emittance growth caused by transverse deflecting fields in a
  linear accelerator}.
\newblock {\em Nucl. Instrum. Meth.}, 178(1):1--8, 1980.

\bibitem{dehler:1998}
M.~Dehler, R.~M. Jones, N.~M. Kroll, R.~H. Miller, I.~Wilson, and W.~Wuensch.
\newblock {Design of a 30 GHz damped detuned accelerating structure}.
\newblock In {\em Proceedings of the 1997 Particle Accelerator Conference (PAC
  97), May 12-16, Vancouver, B.C. Canada}, pages 518--520, 1998.

\bibitem{braun:2008}
H.~Braun et~al.
\newblock {CLIC 2008 Parameters}.
\newblock Technical Report {CERN-OPEN-2008-021, CLIC-Note-764}, CERN, October
  2008.

\bibitem{hoffstaetter:2004}
G.~H. Hoffstaetter and I.~V. Bazarov.
\newblock {Beam-breakup instability theory for energy recovery linacs}.
\newblock {\em Phys. Rev. Spec. Top. - Accel. Beams}, 7(5):054401, 2004.

\bibitem{volkov:2011}
V.~Volkov, J.~Knobloch, and A.~Matveenko.
\newblock {Beam breakup instability suppression in multicell superconducting rf
  guns}.
\newblock {\em Phys. Rev. Spec. Top. - Accel. Beams}, 14(5):054202, 2011.

\bibitem{Egenolf_PRAB_2020}
T.~Egenolf, U.~Niedermayer, and O.~Boine-Frankenheim.
\newblock Tracking with wakefields in dielectric laser acceleration grating
  structures.
\newblock {\em Phys. Rev. ST Accel. Beams}, 23:054402, 2020.

\bibitem{niedermayer_beam_2017}
Uwe Niedermayer, Thilo Egenolf, and Oliver Boine-Frankenheim.
\newblock Beam dynamics analysis of dielectric laser acceleration using a fast
  6d tracking scheme.
\newblock {\em Phys. Rev. Accel. Beams}, 20(11):111302, November 2017.

\bibitem{CST}
{CST Studio Suite}.
\newblock {CST Microwave Studio}, 2008 [http:// www.cst.com].

\bibitem{karagodsky:2006}
V.~Karagodsky, A.~Mizrahi, and L.~Schachter.
\newblock {\em Phys. Rev. ST Accel. Beams}, 9:051301, 2006.

\bibitem{niedermayer:focusing:2018}
U.~Niedermayer, T.~Egenolf, O.~Boine-Frankenheim, and P.~Hommelhoff.
\newblock Alternating phase focusing for dielectric laser acceleration.
\newblock {\em Phys. Rev. Lett.}, 121:214801, 2018.

\bibitem{hanuka:multi:2018}
A.~Hanuka and L.~Schachter.
\newblock {Optimized Operation of Dielectric Laser Accelerators: Multi Bunch}.
\newblock {\em Phys. Rev. Accel. Beams}, 21:064402, 2018.

\bibitem{schachter:kimura:2020}
L.~Schachter and W.~D. Kimura.
\newblock Electron beam guiding by a laser {Bessel} beam.
\newblock {\em Phys. Rev. Accel. Beams}, 23:081301, 2020.

\bibitem{naranjo_stable_2012}
B.~Naranjo, A.~Valloni, S.~Putterman, and J.~B. Rosenzweig.
\newblock Stable {Charged}-{Particle} {Acceleration} and {Focusing} in a
  {Laser} {Accelerator} {Using} {Spatial} {Harmonics}.
\newblock {\em Phys. Rev. Lett.}, 109(16):164803, October 2012.

\bibitem{Niedermayer_PRL_2020}
U.~Niedermayer, T.~Egenolf, and O.~Boine-Frankenheim.
\newblock Three dimensional alternating-phase focusing for dielectric-laser
  electron accelerators.
\newblock {\em Phys. Rev. Lett.}, 125:164801, 2020.

\bibitem{Niedermayer_PhysRevApplied_2021}
U.~Niedermayer, J.~Lautenchl\"{a}ger, Thilo Egenolf, and O.~Boine-Frankenheim.
\newblock Design of a scalable integrated nanophotonic electron accelerator on
  a chip.
\newblock {\em Phys. Rev. Applied}, 16:024022, 2021.

\bibitem{Niedermayer_etal_this_issue}
U.~Niedermayer, K.~Leedle, P.~Musumeci, and S.~A. Schmid.
\newblock Beam dynamics in dielectric laser acceleration.
\newblock {\em J. Instrum. \textit{this issue}}, 2022.

\bibitem{plettner_proposed_2006}
T.~Plettner, P.~P. Lu, and R.~L. Byer.
\newblock Proposed few-optical cycle laser-driven particle accelerator
  structure.
\newblock {\em Phys. Rev. ST Accel. Beams}, 9(11):111301, November 2006.

\bibitem{wei:2017}
Y.~Wei et~al.
\newblock {Dual-gratings with a Bragg reflector for dielectric laser-driven
  accelerators}.
\newblock {\em Phys. Plasmas}, 24(7):073115, 2017.

\bibitem{su:2018}
L.~Su et~al.
\newblock {Fully-automated optimization of grating couplers}.
\newblock {\em Opt. Exp.}, 26(4):4023, 2018.

\bibitem{miller:2015}
D.~A.~B. Miller.
\newblock {Perfect optics with imperfect components}.
\newblock {\em Optica}, 2(8):747--750, 2015.

\bibitem{wang:2015}
J.~Wang.
\newblock {\em {Grating and Ring Based Devices on SOI Platform}}.
\newblock PhD thesis, McGill University, 2015.

\bibitem{chang4}
C.~M. Chang and O.~Solgaard.
\newblock Silicon buried gratings for dielectric laser electron accelerators.
\newblock {\em Appl. Phys. Lett.}, 104(18):184102, 2014.

\bibitem{hughes:avm:2017}
T.~Hughes et~al.
\newblock {Method for computationally efficient design of dielectric laser
  accelerator structures}.
\newblock {\em Opt. Exp.}, 25(13):15414, 2017.

\bibitem{plettner:2008}
T.~Plettner and R.~L. Byer.
\newblock Proposed dielectric-based microstructure laser-driven undulator.
\newblock {\em Phys. Rev. ST Accel. Beams}, 11:030704, 2008.

\bibitem{plettner:2011}
T.~Plettner, R.~L. Byer, and B.~Montazeri.
\newblock {\em J. Mod. Opt.}, 58:1518, 2011.

\bibitem{soong:2012}
K.~Soong, R.~L. Byer, E.~R. Colby, R.~J. England, and E.~A. Peralta.
\newblock Grating-based deflecting, focusing, and diagnostic dielectric laser
  accelerator structures.
\newblock In {\em Proc. of the 2012 Advanced Accelerator Concepts Workshop}.
  A.I.P. Conf. Proc, No. 1507, p. 516, Amer. Inst. Physics, New York, 2012.

\bibitem{soong:2012b}
K.~Soong and R.~L. Byer.
\newblock Design of a subnanometer resolution beam position monitor for
  dielectric laser accelerators.
\newblock {\em Opt. Lett.}, 37:975, 2012.

\bibitem{englandAAC:2018}
R.~J. England, A.~Ody, and Z.~Huang.
\newblock Transverse dynamics in a planar symmetric laser-driven accelerator.
\newblock Technical Report {SLAC-PUB-17451}, SLAC National Accelerator
  Laboratory, 2018.

\bibitem{soong:2014}
K.~Soong, E.~Peralta, R.~J. England, Z.~Wu, E.~R. Colby, I.~Makasyuk, J.~P.
  MacArthur, A.~Ceballos, and R.~L. Byer.
\newblock Electron beam position monitor for a dielectric micro accelerator.
\newblock {\em Opt. Lett.}, 39(16):4747--4750, 2014.

\bibitem{shin:2013}
W.~Shin and S.~Fan.
\newblock {Accelerated solution of the frequency-domain Maxwell's equations by
  enginnering the eigenvalue distribution of the operator}.
\newblock {\em Opt. Express}, 21(19):22578, 2013.

\bibitem{egenolf:2017}
T.~Egenolf, O.~Boine-Frankenheim, and U.~Niedermayer.
\newblock {Simulation of DLA Grating Structures in the Frequency Domain}.
\newblock {\em J. of Phys.: Conf. Ser.}, 874(1):012040, 2017.

\bibitem{niedermayer:2017}
U.~Niedermayer, T.~Egenolf, and O.~Boine-Frankenheim.
\newblock {Beam dynamics analysis of dielectric laser acceleration using a fast
  6D tracking scheme}.
\newblock {\em Phys. Rev. Accel. Beams}, 20(11):111302, 2017.

\bibitem{VSim}
C.~Nieter and J.~R. Cary.
\newblock {VORPAL: a versatile plasma simulation code}.
\newblock {\em J. Comput. Phys.}, 196:448, 2004.

\bibitem{Niedermayer_PRAB_2015}
U.~Niedermayer, O.~Boine-Frankenheim, and H.~De Gersem.
\newblock Space charge and resistive wall impedance computation in the
  frequency domain using the finite element method.
\newblock {\em Phys. Rev. ST Accel. Beams}, 18:032001, 2015.

\bibitem{FLuka}
G.~Battistoni, T.~T. B\"{o}hlen, F.~Cerutti, P.~W. Chin, L.~S. Esposito,
  A.~Fass\`{o}, et~al.
\newblock {Overview of the FLUKA code}.
\newblock {\em Ann. Nucl. Energy}, 82:10--18, 2015.

\end{thebibliography}



\end{document}